\documentclass[twocolumn,prstab,floatfix]{revtex4}
\usepackage{graphicx}
\usepackage{amsmath}
\usepackage{latexsym}
\usepackage{amsfonts}
\usepackage{amssymb}

\begin{document}

\title{Beam-Breakup Instability Theory for Energy Recovery Linacs}
\author{Georg H.~Hoffstaetter} 
\author{Ivan V.~Bazarov}
\affiliation{Laboratory for Elementary Particle Physics,
Cornell University, Ithaca, New York 14853}

\begin{abstract}
Here we will derive the general theory of the beam-breakup instability
in recirculating linear accelerators, in which the bunches do not have
to be at the same RF phase during each recirculation turn.
This is important for the description of energy recovery linacs (ERLs)
where bunches are recirculated at a decelerating phase of the RF wave
and for other recirculator arrangements where different RF phases are
of an advantage.  Furthermore it can be used for the analysis of phase
errors of recirculated bunches.  It is shown how the threshold current
for a given linac can be computed and a remarkable agreement with
tracking data is demonstrated.  The general formulas are then analyzed
for several analytically solvable cases, which show: (a) Why different
higher order modes (HOM) in one cavity do not couple so that the most
dangerous modes can be considered individually. (b) How different HOM
frequencies have to be in order to consider them separately. (c) That
no optics can cause the HOMs of two cavities to cancel. (d) How an
optics can avoid the addition of the instabilities of two
cavities. (e) How a HOM in a multiple-turn recirculator interferes
with itself. Furthermore, a simple method to compute the orbit
deviations produced by cavity misalignments has also been introduced.
It is shown that the BBU instability always occurs before the orbit
excursion becomes very large.
\end{abstract}

\maketitle

\section{Introduction}
Synchrotron light sources based on Energy Recovery Linacs (ERLs) show
promise to deliver X-ray beams with both brilliance and X-ray pulse
duration far superior to the values that can be achieved with storage
ring technology. This is due to the fact that the emittances in an ERL
are largely determined by a laser-driven source, technology which
has been improving steadily over the years, and will undoubtedly
improve further. To generate high brilliance high flux X-rays it is
necessary to accelerate beams
to the energies (several GeV) and with the currents (several $100$\,mA)
that are typical in these storage rings.  This would require that the
linac delivers a power of order $1$\,GW to the beam.  Without somehow
recovering this energy after the beam has been used, such a device
would be practically unfeasible.  Energy recovery \cite{Tigner65_01} can
be achieved by decelerating high energy electrons to generate cavity
fields which in turn accelerate new electrons to high energy.  With
this, large beam powers that are not accessible in a conventional
linac, can be produced.

Several laboratories have proposed high power ERLs for different
purposes.  Designs for light production with different parameter sets
and various applications are being worked on by Cornell University
\cite{CHESS01_03,ERL03_12}, BNL \cite{Benzvi01_02}, Daresbury
\cite{Pool03_01}, TJNAF \cite{Benson01_01}, JAERI
\cite{Sawamura03_02}, the University of Erlangen \cite{Berkaev02_01},
Novosibirsk \cite{Kulipanov98_01}, and KEK \cite{Suwada02_01}. TJNAF
has incorporated an ERL in its design of an electron--ion collider
(EIC) \cite{Merminga02_01} for medium energy physics, while BNL is
working on an ERL--based electron cooler \cite{Benzvi03_01} for the
ions in the relativistic ion collider (RHIC). The work at TJNAF, JAERI
and Novosibirsk is based on existing ERLs of relatively small scale.

One important limitation to the current that can be accelerated in
such an ERL is given by the beam-breakup (BBU) instability.  The size
and cost of all these new accelerators certainly requires a very
detailed understanding of this limitation.

A theory of BBU instability in recirculating linacs, where the energy
is not recovered in the linac, but where energy is added to the beam
when it returns after each recirculation turn, was presented in
\cite{Bisognano87_01}.  Such a linac can consists of many cavities and
several recirculation turns can be used.  This original theory was
additionally restricted to scenarios where the bunches of the
different turns are in the linac at about the same accelerating RF
phase, such as in the so-called continuous wave (CW) operation where
every bucket is filled.  Tracking simulations \cite{Krafft87_01}
compared well with this theory.  In the following we therefore refer
to it as the CW recirculator BBU theory.  It determines above what
threshold current $I_{\rm th}$ the transverse bunch position $x$
displays undamped oscillations in the presence of a higher order mode
(HOM) with frequency $\omega_\lambda$.  If there is only one higher
order mode and one recirculation turn with a recirculation time $t_r$
in the linac, the following formula is obtained for
$T_{12}\sin\omega_\lambda t_r<0$:
\begin{equation}
I_{\rm th} = -\frac{2c^2}{e(\frac{R}{Q})_\lambda Q_\lambda
\omega_\lambda}\frac{1}{T_{12}\sin\omega_\lambda t_r}\ ,
\label{eq:simpbbu}
\end{equation}
where $c$ is the speed of light, $T_{12}$ is the element of the
transport matrix that relates initial transverse momentum $p_x$ before
and $x$ after the recirculation loop, $e$ is the elementary charge,
$(R/Q)_\lambda Q_\lambda$ is the impedance (in units of $\Omega$) of
the higher order mode driving the instability, $Q_\lambda$ is its
quality factor. A corresponding formula had already been presented in
\cite{randbook}. Occasionally, additional factors are found when this
equation is stated \cite{Sereno94,Beard03_01,Merminga01_01}. We give a
concise derivation which shows that no such additional factors are
required.

In the past efforts have been made to derive threshold currents from
the analysis of experimental data obtained at the TJNAF FEL-ERL
\cite{Sereno94,Merminga01_01} using the CW recirculator BBU theory.
The data has not been interpreted satisfactorily, and at least part of
the reason could be that the underlying theory had not been derived
for ERL operation.  The theory presented in this paper should
therefore be used to extend and improve these previous analysis
results, as it is directly applicable to ERL operation.

Here we will derive the general theory of the beam-breakup instability
in recirculating linear accelerators, in which the bunches do not have
to be at the same RF phase during each recirculation turn, similar to
what has first been presented in \cite{Yunn91_01}. First we treat the
simplest case of one dipole HOM and one recirculation turn. Then we
allow many HOMs and many recirculations, and finally we analyze
several analytically solvable cases.

\section{One Dipole HOM and One Recirculation}

For recirculating linacs, the simplest case of one HOM and one
recirculation loop has long been described
\cite{Bisognano86_01}.  
Previous theories which assume that the recirculation time is an integral
number of RF periods should not be used when investigating ERLs. Next we
derive more general formulas that may be applied for arbitrary recirculation
times.

\subsection{The Dispersion Relation}

In the simplest model of multi-pass beam breakup, bunches are
injected into a cavity, which is assumed to have one dipole HOM
(e.g. $TM_{11}$-like mode), accelerated in the cavity and then
recirculated to pass the cavity a second time before they are ejected.
In the case of a two-turn recirculating linac, each bunch would be
accelerated on both passes through the linac and ejected to a user
area.  The RF phase of the bunch would therefore be approximately the
same on both passes.  In an ERL, the RF phase on the second pass
through the cavity is shifted by $\pi$ with respect to the first pass
so that the energy that the bunch gains in the first pass is returned
to the cavity during the second pass and the bunch is ejected with
reduced energy into a beam dump.

If a dipole HOM is excited in the cavity, then a bunch that enters the
cavity on axis experiences a transverse kick and starts to oscillate
around the design orbit of the recirculation loop and returns to the
cavity with a transverse offset.  This offset leads to a change in the
energy of the HOM.  If it increases the HOM energy, transverse kicks
experienced by subsequent bunches will be larger, which will in turn
lead to a further growth of the HOM energy once the kicked bunch
returns to the linac: an instability develops.

To describe this effect, we use the ideas and nomenclature from
\cite{krafft89}. When a current $I(t')$ passes the cavity during its
recovery loop at a time $t'$, the charge $I(t')dt'$ with the transverse
offset $x(t')$ excites the dipole HOM, creating a transverse momentum
for particles traveling through the cavity subsequently at time $t$,
\begin{equation}
\Delta p_x(t)=\frac{e}{c}W(t-t')x(t')I(t')dt'\ ,\label{eq:kick}
\end{equation}
where the wake function $W(\tau)$ describes the transverse force at time
$\tau$ after the HOM was excited.  The momentum transfer is described
by an effective transverse voltage of the HOM, $V(t)= \frac{c}{e} \Delta
p_x(t)$.

Assuming that all bunches are injected on the cavity's central axis,
they do not excite dipole HOMs on their first path through the cavity.
However, the effective transverse voltage of the HOM determines what
kick $\Delta p_x(t)$ the bunch sees and what position it will have
when it returns to the cavity after the recirculation time $t_r$.  The
transfer matrix element $T_{12}$ maps the transverse momentum
$p_x(t)$ to $x(t+t_r)=T_{12} p_x(t)$.  Inserting this into
Eq.~(\ref{eq:kick}) leads to an integral equation for the HOM's
effective voltage,
\begin{equation}
V(t)
= \int_{-\infty}^t W(t-t')I(t')T_{12}\frac{e}{c}V(t'-t_r) dt'\ .
\end{equation}

To solve this integral equation, one assumes that the current is a
continuous stream of short pulses being injected at multiples of an
interval between bunches $t_b$, so that the current on the second turn
is given by
\begin{equation}
I(t)
= I_0 t_b\sum_{m=-\infty}^\infty \delta_D(t - t_r - m t_b)\ ,
\end{equation}
$\delta_D$ being the Dirac-delta function.  Note that $t_b$ is an
integer multiple of the RF circulation time
$t_0=2\pi / \omega_0$ for the RF frequency $\omega_0$.  We write
the recirculation time in terms of the time $t_b$ between bunches as
\begin{equation}
t_r = (n_r - \delta)t_b\ ,
\label{eq:deltadef}
\end{equation}
with an integer $n_r$ and $\delta\in[0,1)$.
For a recirculating linac one has
\begin{equation}
\delta t_b\approx n t_0\ ,
\label{eq:tbdelt}
\end{equation}
and for an ERL one has $\delta t_b\approx (n+\frac{1}{2})t_0$ for some
integer $n$.  A ``$+$'' sign in Eq.~\ref{eq:deltadef} that defines
$\delta$ may seem more natural but our choice leads to simplified
equations.

The HOM voltage at a time $t\in[nt_b+t_r,nt_b+t_r+t_b)$ is given
by
\begin{equation}
V(t)
= I_0 t_b T_{12} \frac{e}{c}\sum_{m=-\infty}^n W(t-t_r-mt_b)V(m t_b)\ .
\label{eq:master}
\end{equation}
Evaluating this at the time $t=nt_b+t_r$ when the recirculated bunches
pass through the cavity leads to
\begin{equation}
V(nt_b+t_r)
= I_0 t_b T_{12} \frac{e}{c}\sum_{m=0}^\infty W(m t_b)V([n-m] t_b)\ .
\end{equation}
In the CW recirculator BBU theory this difference equation was dealt
with by assuming that the voltage can be written as
$V(t)=V_0e^{-i\omega t}$ for $t=nt_b$ where a positive imaginary part
of $\omega$ indicates instability.  Note that this does not require
$V(t)$ to be a harmonic function, but that it can be a linear
combination of harmonic functions with frequencies
$\omega+m\frac{2\pi}{t_b}$ for integers $m$.  This distinction has not
been always made clear and is a potential source of confusion
\cite{Sereno94}.  One obtains the equation
\begin{equation}
\frac{1}{I_0} = t_b T_{12} \frac{e}{c} e^{i\omega t_r}\sum_{m=0}^\infty
W(mt_b)e^{i \omega m t_b}\ .\label{eq:masterold}
\end{equation}
The smallest value of the current $I_0$ for which there is a real
$\omega$ is the threshold current $I_{\rm th}$ of the
instability. 

We proceed by writing $V(t)$ in terms of its
Laplace transform, retaining all possible frequencies in HOM voltage, 
which automatically enables proper description of arbitrary
recirculating configuration:
\begin{equation}
V(t) = \frac{1}{2\pi} \int_{-\infty-ic_0}^{\infty-ic_0} \tilde
V(\omega')e^{-i\omega' t} d\omega'\ .
\label{eq:laplace}
\end{equation}
Note that this is not the conventional way of writing a Laplace
transform. We have chosen this notation in order to make it appear
more similar to a Fourier Transform.  It also makes the subsequent
notation more similar to the CW recirculator BBU theory.  The Laplace
transform is used rather than the Fourier transform since we want to
analyze the onset of instability where the frequencies $\omega$ become
complex.  With the following definition
\begin{equation}
\tilde V^\Sigma(\omega)
=
\sum_{n=-\infty}^\infty \tilde V(\omega+\frac{2\pi}{t_b} n)
\end{equation}
we obtain
\begin{equation}
\tilde V^\Sigma(\omega)
=
t_b\sum_{n=-\infty}^\infty V(n t_b)e^{i \omega n t_b}\ .
\end{equation}
Since $\tilde V^\Sigma(\omega)$ is periodic with $2\pi /t_b$,
it has a Fourier series, and its Fourier coefficients are $V(n t_b)$.
This shows that $\tilde V^\Sigma(\omega)$ does not vanish.  We can
therefore choose $t=(n+n_r)t_b$ in Eq.~(\ref{eq:master}) and sum
over $n$,
\begin{eqnarray}
&&\tilde V^\Sigma(\omega)
=
t_b\sum_{n=-\infty}^\infty V([n+n_r]t_b)e^{i\omega [n+n_r] t_b}\\
&=&
I_0 t_b^2 T_{12}\frac{e}{c}\nonumber\\
&\times&\sum_{n=-\infty}^\infty
\sum_{m=0}^\infty W([m+\delta]t_b)V([n-m]t_b)e^{i\omega [n+n_r] t_b}
\nonumber\\
&=&
I_0 t_b T_{12}\frac{e}{c}e^{i \omega n_r t_b}
\sum_{m=0}^\infty W([m+\delta]t_b)e^{i \omega m t_b}
\tilde V^\Sigma(\omega)\ .
\nonumber
\end{eqnarray}
This finally yields the dispersion relation between $I_0$ and $\omega$
which can be used for all $\delta$.  A corresponding derivation, which has 
treated the beam recirculation in a way that can be applied to ERLs, has
been presented in \cite{Yunn91_01}.  
We believe that this
paper should be referenced more often, since it is hardly referenced,
whereas the earlier papers with CW recirculator BBU theory, which was
not derived for ERLs, are often
referenced in the context of ERLs, where they are not strictly applicable.

\subsection{The Far-Field Wake}

The sum in the dispersion relation
\begin{eqnarray}
\frac{1}{I_0} &=& t_b T_{12} \frac{e}{c} e^{i \omega
n_r t_b}w(\delta)\ ,\label{eq:disp0}
\\ w(\delta) &=& \sum_{n=0}^\infty
W([n+\delta]t_b)e^{i \omega n t_b}\ ,
\end{eqnarray}
can be computed when the far-field approximation for the wake function is
used,
\begin{equation}
W(\tau)=\left(\frac{R}{Q}\right)_\lambda \frac{\omega_\lambda^2}{2c}
e^{-\frac{\omega_\lambda}{2Q_\lambda}\tau}\sin \omega_\lambda\tau \ .
\end{equation}
With $\omega_\lambda^\pm =
\omega_\lambda\pm i\frac{\omega_\lambda}{2Q_\lambda}$ and
$\omega^+=\omega+i\frac{\omega_\lambda}{2Q_\lambda}$ the required sum
can be evaluated if ${\rm
Im}(\omega)>-\frac{\omega_\lambda}{2Q_\lambda}$ and becomes
\begin{eqnarray}
w(\delta)
&=&
\sum_{n=0}^\infty W([n+\delta]t_b)e^{i \omega n t_b}\label{eq:wsum}\\
&=&
\left(\frac{R}{Q}\right)_\lambda\frac{\omega_\lambda^2}{4ic}
\left[
\frac{e^{i\omega_\lambda^+\delta t_b}}{1-e^{i(\omega_\lambda^+ +\omega)t_b}}
-
\frac{e^{-i\omega_\lambda^-\delta t_b}}{1-e^{-i(\omega_\lambda^- -\omega)t_b}}
\right]\nonumber\\
&=&
\left(\frac{R}{Q}\right)_\lambda\frac{\omega_\lambda^2}{4c}
e^{-i\omega\delta t_b}\nonumber\\
&\times&
\frac{
e^{i\omega^+ (\delta-1)t_b}\sin (\omega_\lambda  \delta t_b) -
e^{i \delta   \omega^+ t_b}\sin(\omega_\lambda [\delta-1] t_b)}
{\cos \omega^+ t_b -\cos \omega_\lambda t_b}\ .\nonumber
\end{eqnarray}
The dispersion relation thus becomes
\begin{equation}
I_0
=
\frac{2}{\mathcal{K} T_{12}}e^{-i\omega n_r t_b}
\frac{e^{\frac{\omega_\lambda}{2Q_\lambda}\delta t_b}[
\cos(\omega^+ t_b)-\cos(\omega_\lambda t_b)]
}
{
e^{-i\omega^+t_b}\sin( \delta   \omega_\lambda t_b)-
                 \sin([\delta-1]\omega_\lambda t_b)
}\ ,\label{eq:disp}
\end{equation}
with $\mathcal{K}=t_b
\frac{e}{c^2}(\frac{R}{Q})_\lambda\frac{\omega_\lambda^2}{2}$.  For
$\delta=0$ this becomes the dispersion relation of the CW recirculator
BBU theory:
\begin{equation}
I_0
=
\frac{2}{\mathcal{K}T_{12}}
e^{-i\omega t_r}
\frac{
\cos(\omega^+ t_b)-\cos(\omega_\lambda t_b)
}
{
\sin(\omega_\lambda t_b)
}\ .
\label{eq:resdisp}
\end{equation}
This describes the case when the recirculated bunches are in the same
buckets as the accelerated bunches. When the recirculated bunches are
just between accelerated bunches, then $\delta=\frac{1}{2}$,
\begin{equation}
I_0
=
\frac{1}{\mathcal{K}T_{12}}e^{-i\omega t_r}
\frac{
\cos(\omega^+ t_b)-\cos(\omega_\lambda t_b)
}
{
\cos(\omega^+ t_b/2)\sin(\omega_\lambda t_b/2)
}\ .
\label{eq:erldisp}
\end{equation}
 For the case that every bucket is filled, this would be an ERL with
$t_b=t_0$.  The dispersion relation for ERLs for other $\delta$ is
less simple than Eq.~(\ref{eq:erldisp}) and has $\delta
t_b=(n+\frac{1}{2})t_0$ in Eq.~(\ref{eq:disp}).  This occurs when
the decelerating and accelerating bunches are not perfectly centered
between each other.

\subsection{The Threshold Current}
For a given positive current $I_0$, the values of $\omega$ that
satisfy the dispersion relation Eq.~(\ref{eq:disp}) will in general be
complex.  If they all have negative imaginary parts, the beam motion
is stable.  If one of them has positive imaginary part it will be
unstable.

For small currents the beam motion is stable.  When the current is
increased, at some point, one of these $\omega$ will become real.  At
this point the threshold current is reached.  The threshold current
$I_{\rm th}$ is therefore the smallest current $I_0$ for which there
is a real $\omega$ that satisfies the dispersion relation.  To find
this current, we note that
\begin{equation}
I_0(\omega+\frac{2\pi}{t_b}) = I_0(\omega)\ , \  \
I_0(-\omega^*) = I_0^*(\omega)\ .
\label{eq:period}
\end{equation}
It is therefore sufficient to investigate $\omega\in[0,\pi/t_b]$.

\begin{figure}
\centering
\includegraphics[width=0.8\linewidth,clip]{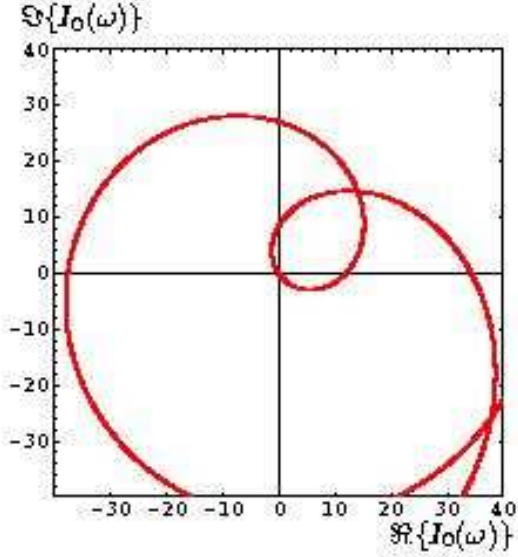}\\
\caption{$I_0(\omega)$ in the complex plain for
$\omega\in[0, \pi /t_b]$. The scale is arbitrary.}
\label{fg:spiral}
\end{figure}

Figure~\ref{fg:spiral} shows $I_0(\omega)$ in the complex plain for
$\omega\in[0,\pi /t_b]$.  The intersection with the real axis
that has the smallest positive value yields the threshold
current.

\section{Approximate Threshold Current}

It can often be justified to linearize in
$\epsilon=\frac{\omega_\lambda}{2Q_\lambda}t_b$, $\epsilon \ll 1$,
which describes a situation when HOM decay is negligible on the time scale 
of the bunch spacing $t_b$. This applies to linacs when nearly every RF bucket
is filled. 
The smallest $|I_0|$ in Eq.~(\ref{eq:disp}) is
obtained when $\cos \omega t_b$ is close to $\cos \omega_\lambda
t_b $, which occurs whenever $\omega$ is close to
$\omega_{\lambda,n_\pm}=\pm \omega_\lambda+n\frac{2\pi}{t_b}$ for any
integer $n$.  Due to Eq.~(\ref{eq:period}), all these frequencies lead
to the same threshold current.  We therefore additionally linearize in
$\Delta\omega=\omega-\omega_{\lambda, n_\pm}$, assuming $\Delta\omega t_b \ll
1$.  This leads to
\begin{equation}
I_0 = \mp\frac{2}{\mathcal{K}T_{12}}e^{-i\omega
n_rt_b}e^{i\omega_{\lambda,0\pm}\delta t_b}(\Delta\omega
t_b+i\epsilon)\ .
\label{eq:klin}
\end{equation}
Within the linearization, the phase factors could be combined to
$e^{i\omega t_r}$.  This is not done in order to retain the symmetries
of Eq.~(\ref{eq:period}) for $\omega\to\omega+2\pi /t_b$ and
for $\omega\to-\omega^*$, i.e.~$\Delta\omega\to-\Delta\omega^*$ and
$\omega_{\lambda,n\pm}\to\omega_{\lambda,-n\mp}$. 

Due to these symmetries, the real current close to
$\omega_{\lambda,n\pm}$ is the same for each of these
frequencies. Without loss of generality we therefore use
$\omega_{\lambda,n\pm}=\omega_\lambda$ and no longer require the symmetries,
\begin{equation}
I_0 = -\frac{2}{\mathcal{K}T_{12}}e^{-i\omega t_r}(\Delta\omega
t_b+i\epsilon)\ .
\label{eq:klin0}
\end{equation}
Since $I_0$ must be real, $\Delta\omega t_b\sin \omega
t_r =\epsilon\cos \omega t_r$ leading to the following two equivalent
equations,
\begin{eqnarray}
I_0 &=& -\frac{\epsilon}{\mathcal{K}}\frac{2}{T_{12}\sin \omega t_r}\
,\\ I_0 &=& -{\rm sign}(\sin \omega t_r)\frac{2}{\mathcal{K}T_{12}}
\sqrt{\epsilon^2+(\Delta\omega t_b)^2}\ .\label{eq:ksquare}
\end{eqnarray}
For this formula to describe the threshold current, it is required
that $I_0>0$ and therefore $T_{12}\sin \omega t_r <0$.

\begin{figure} 
\includegraphics[width=0.48\linewidth, bb=116 616 236 743, clip]{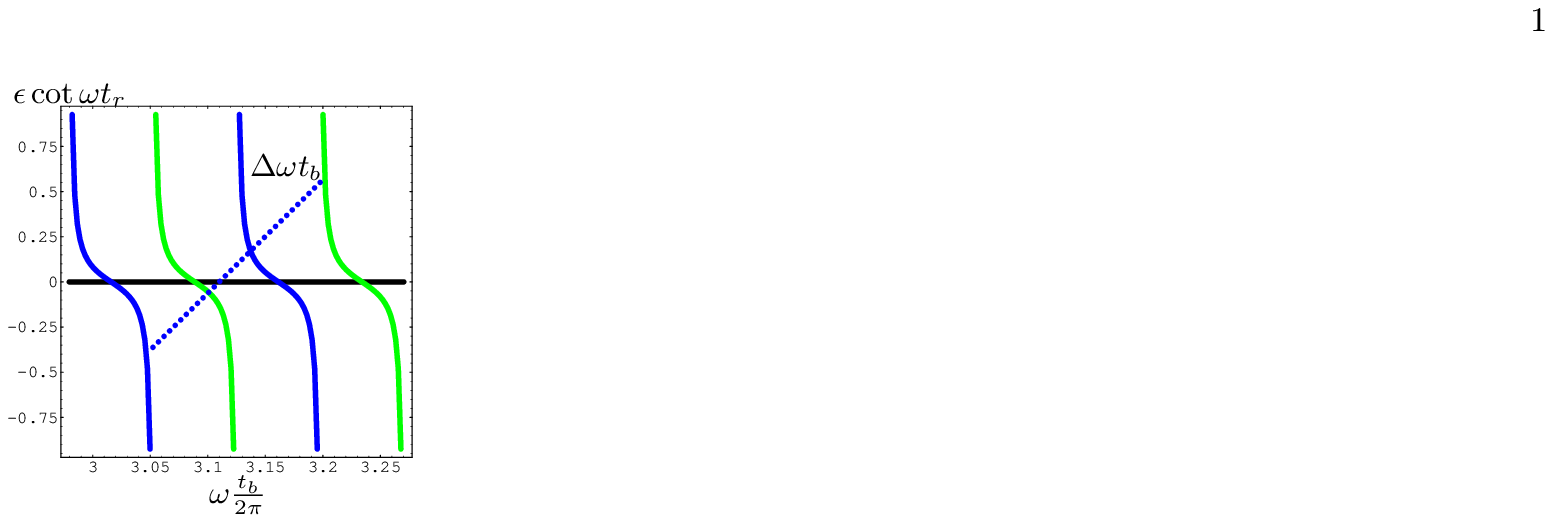}
\includegraphics[width=0.48\linewidth, bb=116 616 236 743, clip]{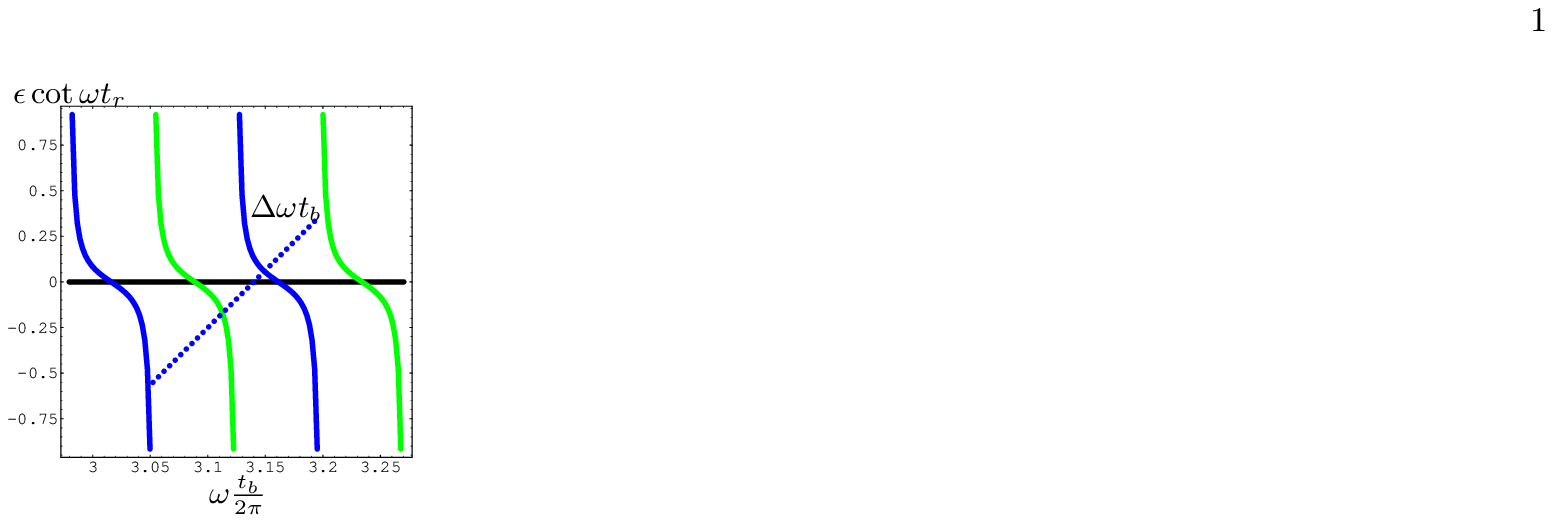}
\caption{$\Delta\omega t_b$ (blue dots) and $\epsilon\cot \omega t_r$
(blue and green curves). The curve is dark blue in the region where
$T_{12}\sin \omega t_r <0$ and light green otherwise. The $\omega$ of
the instability is given by the intersection of the dotted line with a
dark blue solid curve for which $\Delta\omega$ is smallest. Left: A
case of $T_{12}\sin \omega_\lambda t_r <0$. Right: A case of
$T_{12}\sin \omega_\lambda t_r >0$.}
\label{fg:xcotx}
\end{figure}
Figure~\ref{fg:xcotx} shows $\Delta\omega t_b$ and
$\epsilon\cot \omega t_r $ versus $\omega t_b\in[0,\pi]$ for
$t_r=6.88t_b$.  The dotted line and the curve have to meet at a region
where $T_{12}\sin \omega t_t<0$, which is indicated by a dark
blue curve.

When $n_r\epsilon\ll 1$, i.e.~HOM decay is negligible on the time scale of
recirculation time, then $\omega\approx\omega_\lambda$
in the region where $T_{12}\sin \omega_\lambda t_r <0$ and one obtains
\begin{equation}
I_0=-\frac{\epsilon}{\mathcal{K}}\frac{2}{T_{12}\sin \omega_\lambda t_r}\ ,
\label{eq:tradiational}
\end{equation}
which is the traditional and commonly used approximation in
Eq.~(\ref{eq:simpbbu}) which had been derived for $\delta=0$.

In the region with $T_{12}\sin \omega_\lambda t_r>0$ one obtains
$\omega n_r\approx n\pi$, which can be used in Eq.~(\ref{eq:ksquare}),
\begin{equation}
I_0 = \frac{2}{\mathcal{K}|T_{12}|} \sqrt{\epsilon^2+\frac{1}{n_r^2}{\rm
Mod}(\omega_\lambda t_r,\pi)^2}\ .
\end{equation}

For $n_r\epsilon\gg 1$, i.e.~when HOM damping is substantial on the
recirculation time scale, one again uses Eq.~(\ref{eq:ksquare}) with
$\omega t_r\approx (2n\mp\frac{1}{2})\pi$,
where the $+$ or $-$ sign is determined by the sign of
$T_{12}\sin \omega_\lambda t_r$,
\begin{equation}
I_0 = \frac{2}{\mathcal{K}|T_{12}|}\sqrt{\epsilon^2 + \frac{1}{n_r^2}{\rm
Mod}(\omega_\lambda t_r\pm\frac{\pi}{2},2\pi)^2}\ .
\end{equation}
Note that in this case the threshold current weakly depends on $t_r$ and can
be estimated simply by $I_0 = 2\epsilon / \mathcal{K}|T_{12}|$.

One can perform these approximations more accurately, for example, by
approximating $\cot \omega t_r$ by a line
or second order curve.  However, in regions where $n_r\epsilon$ is not
much larger or much smaller than $1$, simple formulas
cannot be found.

\begin{figure} 
\includegraphics[width=0.8\linewidth]{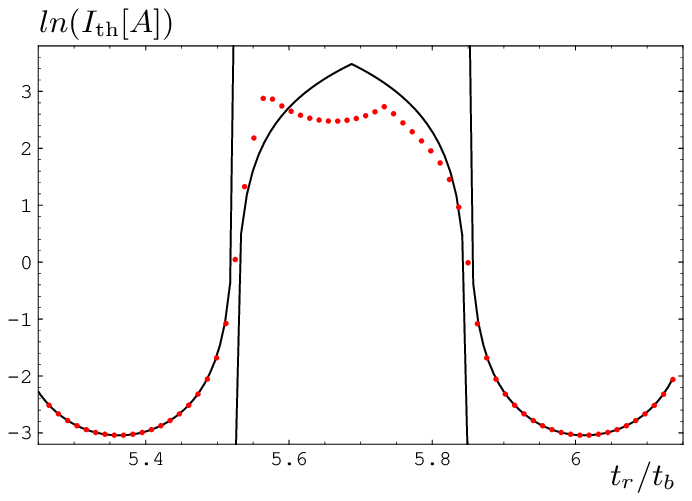}
\includegraphics[width=0.8\linewidth]{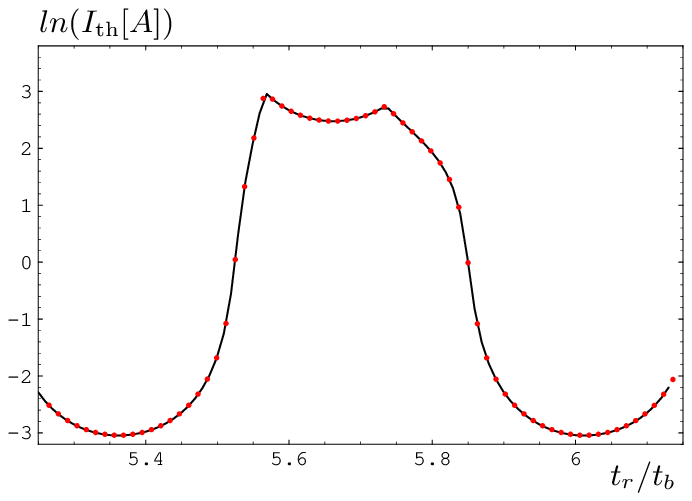}
\caption{Threshold current obtained by tracking (red dots) and
approximate analytical solution (top) and by a numerical solution (bottom) of
the dispersion relation Eq.~(\ref{eq:disp}). Parameters:
$n_r-\delta\in[6.135,7.234]$, $(R/Q)_\lambda=100\,\Omega$,
$Q_\lambda=10^4$, $T_{12}=-10^{-6}$\,eV/c, $\omega_\lambda t_b= 9.67$.}
\label{fg:track}
\end{figure}

Figure~\ref{fg:track}~(top) shows the threshold current obtained with
the approximate analytic solution compared with the threshold current
that is found by tracking particles for the simple case of one cavity
with one HOM and one recirculation loop.

Figure~\ref{fg:track}~(bottom) compares the same tracking results with
a numerical solution of the dispersion relation Eq.~(\ref{eq:disp}).
The data agrees remarkably well with the approximate formula in the
region where $\Delta\omega$ is small, i.e. where $I_0$ is relatively
small. In the region where the threshold current is relatively large,
the agreement with the approximate expression is not satisfactory, however.

To find the threshold current with Eq.~(\ref{eq:disp}), the smallest
positive real value of $I_0$ for $\omega\in[0,\pi /t_b]$ was
found by linearly interpolating 100 points in the region
$\omega\in[\omega_\lambda-\frac{1}{n_r}\frac{\pi}{t_b} ,
\omega_\lambda+\frac{1}{n_r}\frac{\pi}{t_b}]$.

\subsection{Instability Growth Rate}

We denote the threshold current by $\hat I_0$ and the real frequency
$\hat \omega\in [0,\pi / t_b]$ satisfies Eq.~(\ref{eq:disp}) for
this current.  When the current $I_0$ is slightly larger than $\hat
I_0$, there is one frequency $\omega(I_0)$ that satisfies
Eq.~(\ref{eq:disp}) and is close to $\hat\omega$.  It has a
positive imaginary part.  All other frequencies $\omega$ at which Eq.~(\ref{eq:disp}) holds and for which therefore $\tilde
V^\Sigma(\omega)$ might not vanish have an imaginary part that is not
positive.  In Eq.~(\ref{eq:laplace}) there are therefore
exponentially growing terms.  For currents that are only slightly
larger than the threshold current, the complex frequency $\omega(I_0)$
can be expanded with respect to $\Delta I=I_0-\hat I_0$,
\begin{equation}
e^{-i\omega t}
=
e^{-i[\hat\omega + {\rm \Re}\{\frac{d\omega}{dI_0}\Delta I\}]t}
e^{{\rm \Im}\{\frac{d\omega}{dI_0}\Delta I\}t}\ .
\end{equation}
The rise time per current of the instability is thus given by
$\alpha={\rm \Im}\{\frac{d\omega}{dI_0}\}|_{\hat I_0}$.  The dispersion
relation Eq.~(\ref{eq:disp}) leads to a long formula.  However, using
the simplified Eq.~(\ref{eq:klin}) leads to
\begin{eqnarray}
\alpha &=& \left. \Im \big\{ \frac{1}{dI_0/d\omega}\big\}\right|_{\hat I_0}
\label{eq:rate}\\
&=& \frac{1}{\hat I_0} \Im
\big\{ \big(-in_r t_b +
\frac{t_b}{(\hat\omega-\omega_\lambda)t_b + i\epsilon}\big)^{-1} 
\big\} \
\nonumber\\ 
&=& \frac{1}{\hat I_0t_b}\frac{4\epsilon +
(\hat I_0 \mathcal{K} T_{12})^2n_r}{4+8n_r\epsilon+
(\hat I_0 \mathcal{K} T_{12}n_r)^2}\ .
\nonumber
\end{eqnarray}
Provided parameters are not in the region where the curves diverge in
Fig.~\ref{fg:track}~(top), i.e. $\sin \omega_\lambda t_r$ is not close
to zero, the following approximate formulas hold for the growth rate:
For $n_r\epsilon\ll 1$ one obtains $\alpha=\frac{1}{\hat I_0
t_b}\epsilon$ and for $n_r\epsilon\gg 1$ one obtains
$\alpha=\frac{1}{\hat I_0 t_b}\frac{1}{n_r}$.

\section{Multiple Dipole HOMs and Multiple Recirculations}

Recirculating linacs with many cavities and several recirculation
loops have been considered early on \cite{Bisognano87_01, krafft89}.
Here we use
the same nomenclature as much as possible.  The $N$ higher order
modes, which can be associated with different cavities, are numbered
by an index $i$.  The $N_p$ passes through the linac are numbered by
an index $I$.  The horizontal position and momentum that the beam has
at time $t$ in the HOM $i$ during turn $I$ is denoted $\vec
z_i^I(t)=(x_i^I(t),p_i^I(t))$.  The transport matrix that transports
the phase space vector $\vec z_j^J$ at HOM $j$ during turn $J$ to
$\vec z_i^I$ is denoted ${\bf T}_{ij}^{I\!J}$ and the time it takes
to transport a particle from the beginning of the first turn to HOM
$i$ during turn $I$ is denoted $t_i^I$.  The beam is propagated from
after HOM $i-1$ to after HOM $i$ by
\begin{equation}
\vec z_i^I(t) = {\bf T}_{ii-1}^{II} \cdot \vec
z_{i-1}^I(t-[t_{i}^I-t_{i-1}^I])+\left(\begin{array}{c}0\\ \frac{e}{c}
V_i(t)\end{array}\right)\ .
\end{equation}
This equation can be iterated to obtain the phase space coordinates as
a function of the HOM strength that creates the orbit oscillations.
With the matrix element $T_{ij}^{I\!J}=({\bf T}_{ij}^{I\!J})_{12}$
one obtains
\begin{eqnarray}
x_i^I(t)
&=&
\sum_{J=1}^{I-1}\sum_{j=1}^N T_{ij}^{I\!J} \frac{e}{c}
V_j(t-[t_i^I-t_j^J])\label{eq:xfromv}\\
&+&
\phantom{\sum_{J=1}^{I-1}}
\sum_{j=1}^{i-1} T_{ij}^{II} \frac{e}{c}
V_j(t-[t_i^I-t_j^I])\nonumber\ .
\end{eqnarray}

The strength $V_i(t)$ of the HOM $i$ is created by all particles that have
traveled through that HOM via the integral
\begin{equation}
V_i(t)=\int_{-\infty}^t \sum_{I=1}^{N_p}W_i(t-t')I_i^I(t')x_i^I(t')dt'\ ,
\label{eq:homsum}
\end{equation}
where $I_i^I(t)$ is the current at time $t$ that the fraction of the
beam has which passes the HOM $i$ on turn $I$.  Combining this with
Eq.~(\ref{eq:xfromv}) leads to the following integral-difference
equation:
\begin{eqnarray}
V_i(t)&=&\int_{-\infty}^t\sum_{I=1}^{N_p} W_i(t-t')I_i^I(t')\\
&\times&\frac{e}{c} \sum_{J=1}^I\sum_{j=1}^{N_{I\!J}(i-1)}T_{ij}^{I\!J}
V_j(t'-[t_i^I-t_j^J])dt'\ ,\nonumber\\
N_{I\!J}(i-1)&=&
\begin{cases}
  N, &\text{if $I\ne J$;} \\
  i-1, &\text{if $I = J$.}
\end{cases}
\end{eqnarray}

Now the approximation of short bunches is used.  The current is given
at time $t$ by pulses that are equally spaced with the distance $t_b$,
\begin{equation}
I_i^I(t)=\sum_{m=-\infty}^\infty I_0 t_b \delta(t-t_i^I-m t_b)\ .
\end{equation}
This reduces the integral to a sum,
\begin{eqnarray}
V_i(t) &=& \frac{e}{c}I_0 t_b \sum_{m=-\infty}^{n(t,t_i^I)}
\sum_{I=1}^{N_p} W_i(t-t_i^I-m t_b)\label{eq:hommultsum}\\ &\times&
\sum_{J=1}^I\sum_{j=1}^{N_{I\!J}(i-1)} T_{ij}^{I\!J}V_j(m
t_b+t_j^J)\ ,\nonumber
\end{eqnarray}
where $n(t,t_i^I) = {\rm Max}_m(t\ge mt_b+t_i^I)$. Computing
\begin{equation}
V_i^L=\sum_{n=-\infty}^\infty V_i(nt_b+t_i^L)e^{i\omega nt_b}
\end{equation}
leads to
\begin{eqnarray}
V_i^L &=& \frac{e}{c}I_0 t_b \sum_{n=-\infty}^\infty
\sum_{m=m^*}^\infty
\sum_{I=1}^{N_p} W_i(m t_b+t_i^L-t_i^I)\\ &\times&
\sum_{J=1}^I\sum_{j=1}^{N_{I\!J}(i-1)} T_{ij}^{I\!J}V_j([n-m]
t_b+t_j^J)e^{i\omega nt_b}\ .\nonumber
\end{eqnarray}
The second summation starts at $m^*=n-n(nt_b+t_i^L,t_i^I)$ which can
be simplified by writing $t_i^I=(n_i^I-\delta_i^I)t_b$,
\begin{eqnarray}
m^*
&=&
n-n([n+n_i^L-n_i^I] t_b,[\delta_i^-\delta_i^I]t_b)\\
&=&
-{\rm Max}_m([n_i^L-n_i^I]t_b\ge [m+\delta_i^L-\delta_i^I]t_b)
\nonumber\\
&=&
n_i^I-n_i^L+\gamma_i(I,L)\ ,\nonumber
\end{eqnarray}
where $\gamma_i(I,L)=1$ if $\delta_i^L > \delta_i^I$ and $0$
otherwise.  Shifting the summation index $m$ now leads to
\begin{eqnarray}
V_i^L
&=&
\frac{e}{c}I_0 t_b \sum_{n=-\infty}^\infty
\sum_{m=0}^\infty \; \: \sum_{I=1}^{N_p}\\
&& W_i([m+\gamma_i(I,L)+\delta_i^I-\delta_i^L] t_b)
\sum_{J=1}^I\sum_{j=1}^{N_{I\!J}(i-1)}T_{ij}^{I\!J}
\nonumber\\
&\times&
V_j([n-m-n_i^I+n_i^L-\gamma_i(I,L)] t_b+t_j^J)e^{i\omega
nt_b}\ .\nonumber
\end{eqnarray}
and with $\delta_i(I,L)=\gamma_i(I,L)+\delta_i^I-\delta_i^L$, which is
between $0$ and $1$, shifting the index $n$ finally leads to the
relation
\begin{eqnarray}
V_i^L &=& \frac{e}{c}I_0 t_b 
\sum_{m=0}^\infty
\sum_{I=1}^{N_p} W_i([m+\delta_i(I,L)] t_b)
\label{eq:viksum}\\
&\times&
\sum_{J=1}^I\sum_{j=1}^{N_{I\!J}(i-1)} T_{ij}^{I\!J}
e^{i\omega [m+n_i^I-n_i^L+\gamma_i(I,L)]t_b}V_j^J\ .\nonumber
\end{eqnarray}

The following sum is equivalent to that in
Eq.~(\ref{eq:wsum}),
\begin{equation}
w_i(\delta) = \sum_{m=0}^{\infty}
W_i([m+\delta]t_b)e^{i\omega m t_b}\ .
\end{equation}
Equation (\ref{eq:viksum}) reduces to
\begin{eqnarray}
\frac{1}{I_0}V_i^L
&=&
\frac{e}{c}t_b\sum_{I=1}^{N_p}w_i(\delta_i(I,L))
e^{i\omega(t_i^I-t_i^L+\delta_i(I,L)t_b)}\nonumber\\
&\times&
\sum_{J=1}^I\sum_{j=1}^{N_{I\!J}(i-1)} T_{ij}^{I\!J}V_j^J\ .
\label{eq:matdef}
\end{eqnarray}
If a vector $\vec V$ is introduced that has the coefficients $V_i^I$,
this equation can be written in matrix form,
\begin{equation}
\frac{1}{I_0}\vec V={\bf W}(\omega){\bf U} \vec V\ ,
\label{eq:vmat}
\end{equation}
where the matrix ${\bf M}={\bf W}(\omega){\bf U}$ is
determined by Eq.~(\ref{eq:matdef}).  When all electrons are
considered to have the speed of light, $t_i^I-t_i^L$ does not depend
on the HOM number $i$ and we therefore drop this index and obtain the
following matrix coefficients:
\begin{eqnarray}
M_{ij}^{LJ} &=&
\frac{e}{c}t_b\sum_{I=J+\Theta_{j,i}}^{N_p}w_i(\delta(I,L))
e^{i\omega{\rm Top}(\frac{t^I-t^L}{t_b})t_b} T_{ij}^{I\!J}\nonumber\\
\Theta_{j,i}&=&
\begin{cases}
  1, &\text{if $j \ge i$;} \\
  0, &\text{otherwise,}
\end{cases}
\label{eq:mmat}
\end{eqnarray}
where ${\rm Top}(x)$ is the smallest integer that is equal to or
larger than $x$.  With Kronecker $\hat\delta_{ik}$ this determines the
matrices ${\bf W}$ and ${\bf U}$ to be
\begin{eqnarray}
W_{ik}^{LI} &=&\frac{e}{c}t_b w_i(\delta(I,L))
e^{i\omega{\rm Top}(\frac{t^I-t^L}{t_b})t_b} \delta_{ik}\ ,\label{eq:wdef}\\
U_{kj}^{I\!J} &=&T_{kj}^{I\!J}\Theta_{I,J+\Theta_{j,k}}\ .\label{eq:udef}
\end{eqnarray}

For each frequency $\omega$, $I_0^{-1}$ is an eigenvalue of
${\bf M}(\omega)$.  Since the eigenvalues are in general complex,
but $I_0$ has to be real, the threshold current is determined by the
largest real eigenvalue of ${\bf M}(\omega)$.  The matrix has the
properties
\begin{equation}
{\bf M}(\omega+\frac{2\pi}{t_b})={\bf M}(\omega)\ ,\ \ 
{\bf M}(-\omega^*)={\bf M}^*(\omega)\ ,
\end{equation}
and it is therefore again sufficient to investigate
$\omega\in[0,\pi /t_b]$ to find the threshold current.

Note that $V_N^{N_p}$ never appears on the right hand side of
Eq.~(\ref{eq:viksum}) so that the dimension of ${\bf M}$ can be
reduced by one to $N\times N_p-1$.  Furthermore the dimension can be
reduced when two fractional parts $\delta_i^I$ and $\delta_j^J$ are
equal since then $V_i^I$ and $V_j^J$ are identical.
Note also that for $N=1$ and $N_p=2$ Eq.~(\ref{eq:viksum}) reduces to
the dispersion relation for one HOM in Eq.~(\ref{eq:disp}).

\subsection{Instability Growth Rate}
The growth rate of the instability is again computed by first
obtaining the threshold current $\hat I_0$ and the real frequency
$\hat\omega$ for which $\hat I_0^{-1}$ is an eigenvalue of ${\bf
M}(\hat\omega)$.  If this is the $k$th eigenvalue $\lambda_k(\omega)$
of the matrix ${\bf M}(\omega)$, then the growth rate of the
instability is given by
\begin{equation}
\alpha=-{\hat I}_0^{-2}\left. \Im
\big\{\big(\frac{d\lambda_k}{d\omega}\big)^{-1}\big\}\right|_{I_0=\hat I_0}\ .
\end{equation}

\section{Multiple HOMs in one Cavity}
The presented theory for multiple HOMs and multiple recirculation
turns can in general only be evaluated with computers.  However, for
some simple situation an analytical understanding is possible.

One such case is an accelerator with one recirculation loop and one
cavity in which many HOMs can be excited. The $(1,2)$ matrix elements
that refer to transport between HOMs for the same pass are zero,
$T^{J\!J}_{ij}=0$.  All matrix elements that describe the recirculation
loop are identical, $T^{I\!J}_{ij}=T_{12}$ if $I\ne J$.  The matrix
elements in Eq.~(\ref{eq:mmat}) are then given by
\begin{equation}
M^{L2}_{ij} = 0\ ,\\
M^{L1}_{ij} = \frac{e}{c}t_bw_i(\delta(2,L))e^{i\omega{\rm
Top}(\frac{t^2-t^L}{t_b}t_b)}T_{12}\ .
\end{equation}
Equation (\ref{eq:vmat}) becomes with $\delta=\delta(2,1)$,
\begin{equation}
\frac{1}{I_0}V^1_i = \frac{e}{c}t_b \sum_j w_i(\delta)e^{i\omega n_r
t_b}T_{12} V^1_j\ .
\end{equation}
A summation over the index $i$ shows that 
$\sum_j V^1_j$ can only be nonzero when
\begin{equation}
\frac{1}{I_0} = t_bT_{12}\frac{e}{c}e^{i\omega n_r t_b}\sum_i
w_i(\delta)\ .
\label{eq:i0multhom}
\end{equation}
Comparing this with Eq.~(\ref{eq:disp0}) shows that one only has to
replace $w(\delta)$ of the single HOM by $\sum_i w_i(\delta)$ to
arrive at the threshold formula for the multi-HOM case.  To find the
smallest real $I_0$ that Eq.~(\ref{eq:i0multhom}) can produce for real
$\omega$, $\sum_i w_i(\delta)$ has to be maximized.  The sum is
especially large when the denominator of one of the terms is very
small, i.e. when $\cos \omega t_b \approx \cos \omega_{\lambda} t_b$.

When the HOM frequencies modulo $2\pi /t_b$
are sufficiently different for the different HOMs, the maximal
absolute value of $\sum_i w_i(\delta)$ will be close to the
largest absolute value that any of the $w_i(\delta)$ could have
individually.  This is due to the fact that all the $w_j(\delta)$ for
$j\ne i$ are relatively small for frequencies $\omega$ for which the
denominator of $w_i(\delta)$ is small.

For HOM frequencies that are sufficiently different in the above
sense, the threshold current for several HOMs therefore does not
differ significantly from the threshold current of the worst
individual HOM.

\begin{figure}
\includegraphics[width=\linewidth]{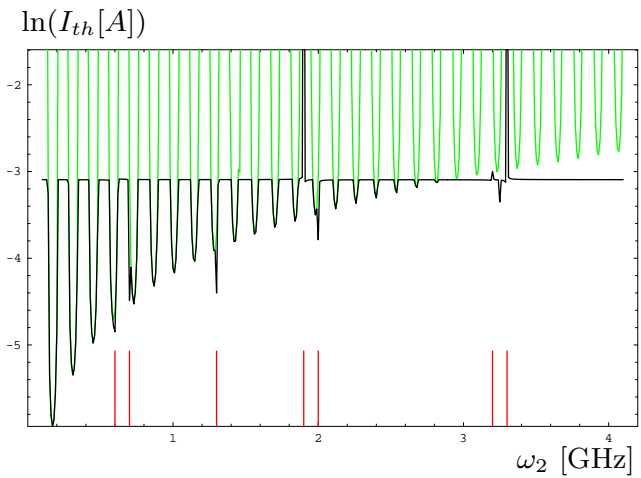}
\caption{Dark black curve: the threshold current $\ln(I_{th}{\rm [A]})$
for one HOM at $\omega_1/2\pi=2\,$GHz as a function of a second HOM with
frequency $\omega_2$. Light green curve: threshold current
when only the second HOM is present. Light red lines:
frequencies for which $\cos \omega_2 t_b \approx \cos \omega_1 t_b$
where the threshold current is not simply the minimum of the threshold
currents produced by the individual HOMs.}
\label{fg:twohom}
\end{figure}

Figure~\ref{fg:twohom} shows how the threshold current changes when
the frequency of one HOM is fixed at a small threshold current with
$|\sin \omega_1 t_r|=1$ and a second HOM frequency is varied.
Superimposed is $I_{th}$ if only the second HOM is present.  It is
apparent that the HOM that would produce the larger threshold-current
if it was solely present only influences the threshold current of the
pair when the frequencies $\pm\omega_1 \, {\rm mod} \, 2\pi /t_b$ and
$\pm\omega_2 \,{\rm mod} \, 2\pi /t_b$ are closer together than
about
$\Delta\omega_\lambda=\epsilon /t_b = \omega_\lambda /2Q_\lambda$.

One can draw the conclusion that in the case of many HOMs they do not
interact destructively when the frequencies $\pm\omega_\lambda \,
{\rm mod} \, 2\pi /t_b$ are not very close together.  Tracking
simulations also demonstrate this effect.

\begin{figure}[ht!]
\begin{minipage}{\linewidth}
\includegraphics[width=\linewidth]{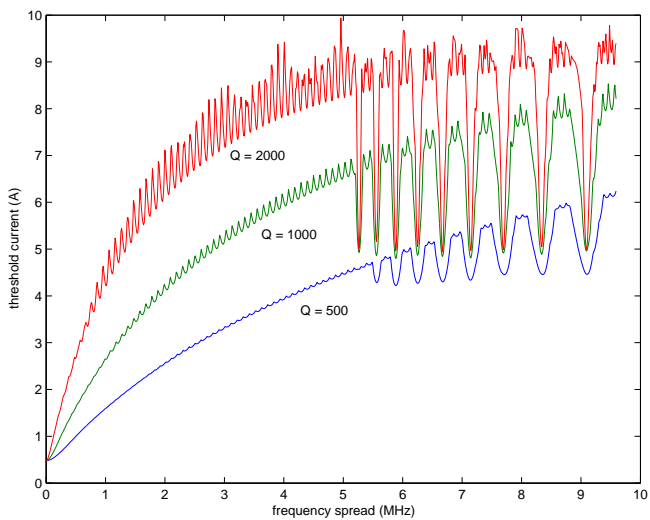}
\end{minipage}
\caption{The threshold current $I_{th}$ as a function of uniform equidistant
frequency spread for 20 HOMs in a single cavity. Abscissa displays
frequency difference between the two HOMs adjacent in frequency.
Parameters: $\bar\omega_\lambda/2\pi = 2\,$GHz,
$(R/Q)_\lambda Q_\lambda = 5000\,\Omega$, $T_{12} = -10^{-6}\,$eV/c,
$t_r = 1.000125\cdot 10^{-6}\,$s, $\omega_0/2\pi = 1.3\,$GHz.
Threshold is determined by tracking with accuracy $0.1\%$.
}
\label{fg:homspread}
\end{figure}

One strategy to increase the BBU threshold current is the introduction
of HOM frequency spreads between cavities.  As an example,
Fig.~\ref{fg:homspread} shows the threshold current found by tracking
as a function of uniform frequency spread of 20 HOMs in a single
cavity. For all three curves $(R/Q)_\lambda Q_\lambda$ is the same. It
is seen that for lower $Q_\lambda$, the curve begins to saturate for
larger frequency spread than for the high $Q_\lambda$ case. It is also
seen that the threshold for frequency spread $\omega_\lambda /
2Q_\lambda$ is similar in all three cases.  Both observations are
consistent with the above assertion that HOMs do not interfere when
they are further apart than $\omega_\lambda / 2Q_\lambda$.  The fact
that oscillations in the Fig.~\ref{fg:homspread} are smaller for low
$Q_\lambda$ is consistent with the conclusion that a broader HOM
resonance peak should lead to more overlaps and as a result to less
pronounced differences in the threshold for different HOM frequencies.

\section{One Dipole HOM in Two Cavities}
In order to see how two dipole HOMs interact, we will now analyze a
set of two HOMs with one recirculation loop, i.e.~$N_p=2$.  We
abbreviate $\delta=\delta(2,1)$ and $\alpha_r=e^{i\omega n_r t_b}$.  The
dispersion relation is then given by
\begin{equation}
\frac{1}{I_0}
\left(
\begin{array}{c}
V_1^1 \\
V_2^1 \\
V_1^2
\end{array}
\right)
=
\frac{e}{c}t_b
{\bf N}
\left(
\begin{array}{c}
V_1^1 \\
V_2^1 \\
V_1^2
\end{array}
\right)
\end{equation}
where the matrix ${\bf N}$ is given by
\begin{equation}
\left(
\begin{array}{ccc}
w_1(\delta)\alpha_r T_{11}^{21}
&
w_1(\delta)\alpha_r T_{12}^{21}
&
0
\\
w_2(0)T_{21}^{11}+w_2(\delta)\alpha_r T_{21}^{21}
&
w_2(\delta)\alpha_r T_{22}^{21}
&
w_2(\delta)\alpha_r T_{21}^{22}
\\
w_1(0)T_{11}^{21}
&
w_1(0)T_{12}^{21}
&
0
\end{array}
\right)\ .
\end{equation}
The third row is similar to the first row and
eliminating it by similarity transformations
leads to the $2\times 2$ matrix
\begin{eqnarray}
&&\left(
\begin{array}{cc}
w_1(\delta)\alpha_r&0\\
0&w_2(\delta)\alpha_r
\end{array}
\right)\times
\label{eq:matn}
\\
&&
\left(
\begin{array}{ccc}
T_{11}^{21}
&
T_{12}^{21}
\\
\frac{w_2(0)}{w_2(\delta)\alpha_r}T_{21}^{11}+T_{21}^{21}
+\frac{w_1(0)}{w_1(\delta)\alpha_r} T_{21}^{22}
&
T_{22}^{21}
\\
\end{array}
\right)\ .\nonumber
\end{eqnarray}

Even though $\omega_1(\delta)$ appears in the denominator, the formula
for the eigenvalues of this matrix does not contain such a
denominator.  Therefore the largest real eigenvalue will again occur
at a frequency $\omega$ for which one of the HOM frequencies satisfies
$\cos(\omega t_b)\approx\cos(\omega_\lambda t_b)$. The term
$w_i(\delta)$ and $w_i(0)$ of the other HOM can then again be
neglected, so that for sufficiently different HOM frequencies
${\rm mod}\:2\pi /t_b$ the threshold current is again approximately
determined by the HOM which would have the smallest $I_{th}$ if there
were no other HOMs present.

We therefore now assume that the two HOMs are equal, $w_1=w_2$.  Now
one can perform an approximation analogous to Eq.~(\ref{eq:klin0})
leading to
\begin{eqnarray}
&& {\bf N} = -\frac{\mathcal{K}}{2}e^{i\omega t_r}
\frac{1}{\Delta\omega t_b+i\epsilon} \times
\label{eq:matlin0}
\\
&&
\left(\begin{array}{cc}
T_{11}^{21}        & T_{12}^{21}\\
T_{21}^{21} +
e^{-i\omega t_r}(T_{21}^{11}+T_{21}^{22}) & T_{22}^{21}
\end{array}\right)\ .
\nonumber
\end{eqnarray}

\begin{figure}[b!]
\begin{minipage}{\linewidth}
\includegraphics[width=\linewidth,clip]{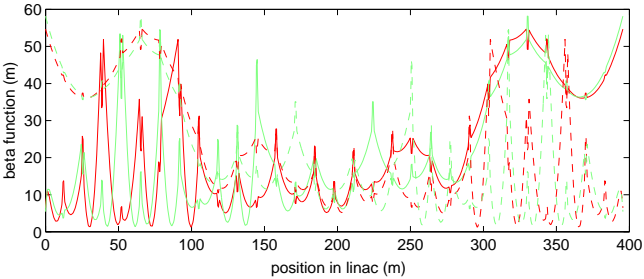}
\end{minipage}
\caption{Example of a mirror-symmetric linac optics. Dark red curve (light
green curve) is horizontal (vertical) beta function in the linac for
accelerating beam. Dashed lines show lattice function
for decelerating beam.}
\label{fg:linacsym}
\end{figure}

It is interesting to analyze whether the effect of the two HOMs can
cancel. A cancellation could occur most naturally when the linac
and the recovery loop are mirror symmetric.  A mirror symmetry of the
linac means that the beta functions of the first pass, going from low
to high energy, are the mirror image of those of the second pass,
going from high to low energy. An example of such an optics is shown
in Fig.~\ref{fg:linacsym}.  

Since the $(1,2)$ element of the transport matrix between a region
with momentum $p_0$ and a region with momentum $p$ can be written with
Twiss parameters as
\begin{equation}
T_{12} = \sqrt{\frac{\beta \beta_0}{p p_0}}\sin\Delta\Psi\ ,
\end{equation}
this symmetry leads to $T_{21}^{22}=T_{21}^{11}$. An additional mirror
symmetry of the return arc leads to $T_{22}^{21}=T_{11}^{21}$. The
eigenvalues of the matrix in Eq.~(\ref{eq:matlin0}) then become,
\begin{eqnarray}
\frac{1}{I_0} &=& -\frac{\mathcal{K}}{2}e^{i\omega t_r}\frac{1}{\Delta\omega
t_b+i\epsilon}\times\\ &&\left[T_{11}^{21}
\pm\sqrt{T_{12}^{21}(T_{21}^{21} + 2e^{-i\omega
t_r}T_{21}^{11})}\right]\ . \nonumber
\end{eqnarray}
Since there are two solutions to the quadratic eigenvalue equation, to
every eigenvalue that is smaller than $1/I_{\rm th}$ of a single
cavity, there exists in general the one that is larger.  Therefore two
cavities do not compensate their instabilities, but it is possible to
decouple the cavities to the extent that the combined threshold
current is just as large as that for a single cavity.  For this, one
has to choose $T_{12}^{21}=0$, i.e.~the phase advance of the return
arc has to be a multiple of $\pi$.

Since the kick of a HOM disturbs the beam most at low energy, the
first and the last cavity of an ERL are the strongest contributors to
BBU. It seems therefore advisable to adjust the phase advance of the
arc to a multiple of $\pi$ also when the linac has more than two
cavities.

\subsection{Multiple Recirculation Turns}

One could envision a multi-turn recirculating linac as ERL.  The beam
would pass the same linac $N_r$ times to reach its top energy, and
subsequently it would be decelerated in just as many turns through the
linac.  The current in the cavities would be $2N_r$ times higher than
the current that is available at high energy.  One could therefore
conjecture that the BBU threshold current is $N_r$ times smaller than
for a one-turn ERL. Here we will show that the threshold current can be
significantly smaller than that conjecture and in general can be expected 
to decrease
quadratically with $N_r$.

Each bunch passes the linac $N_p=2N_r$
times and the matrix in Eq.~(\ref{eq:mmat}) becomes
\begin{equation}
M^{LJ}=\frac{e}{c}t_b\sum_{I=J+1}^{N_p}w(\delta(I,L))e^{i\omega{\rm
Top}(\frac{t^I-t^L}{t_b})t_b}T^{I\!J}\ ,
\end{equation}
where the lower indexes have been suppressed since only one HOM is
considered.  The eigenvalue equation thus becomes
\begin{equation}
\frac{1}{I_0}V^L
=
\frac{e}{c}t_b\sum_{J=1}^{N_p}\sum_{I=J+1}^{N_p}
w(\delta(I,L))e^{i\omega{\rm Top}(\frac{t^I-t^L}{t_b})t_b}T^{I\!J}V^J\ ,
\end{equation}

An approximation equivalent to that leading to Eqs.~(\ref{eq:klin0})
and (\ref{eq:matlin0}) results in
\begin{eqnarray}
&&\frac{1}{I_0}V^L
=
-\frac{\mathcal{K}}{2}\frac{1}{\Delta\omega t_b+i\epsilon}\times\\
&&\sum_{J=1}^{N_p}\sum_{I=J+1}^{N_p}
e^{i\omega[{\rm Top}(\frac{t^I-t^L}{t_b})-\delta(I,L)]t_b}T^{I\!J}V^J\ .
\nonumber
\end{eqnarray}
Comparing the exponent on the right hand side to those in
Eqs.~(\ref{eq:matdef}) and (\ref{eq:mmat}) shows that this can be written as
\begin{eqnarray}
&&\frac{1}{I_0} V^L e^{i\omega t^L}
=
-\frac{\mathcal{K}}{2}\frac{1}{\Delta\omega t_b+i\epsilon}\times\\
&&\sum_{J=1}^{N_p}\sum_{I=J+1}^{N_p}
e^{i\omega(t^I-t^J)}T^{I\!J}V^J e^{i\omega t^J}\ .
\nonumber
\end{eqnarray}
Since the right hand side does not depend on $L$, all the terms $V^J
e^{i\omega t^J}$ are equivalent and the condition that they do not
vanish is
\begin{equation}
\frac{1}{I_0}
=
-\frac{\mathcal{K}}{2}\frac{1}{\Delta\omega t_b+i\epsilon}
\sum\!\!\!\!\!\!\!\!\sum\limits_{I>J}
e^{i\omega(t^I-t^J)}T^{I\!J}\ ,
\end{equation}
where the double 
sum goes over all pairs of $I$ and $J$ for which
$I>J$.  Similar to the condition obtained from Eq.~(\ref{eq:klin0})
which is analyzed with Fig.~\ref{fg:xcotx}, the fact that $I_0$ has
to be real entails the condition
\begin{equation}
\frac{
\sum\!\!\!\!\!\!\!\!\sum\limits_{I>J} \cos(\omega[t^I-t^J])T^{I\!J}
}{
\sum\!\!\!\!\!\!\!\!\sum\limits_{I>J} \sin(\omega[t^I-t^J])T^{I\!J}
}
=\frac{\Delta\omega}{\epsilon}\ .
\end{equation}
In regions where $\sum\!\!\!\!\!\!\!\!\sum_{I>J}
\sin(\omega[t^I-t^J])T^{I\!J}>0$, we again obtain the approximation that
$\omega\approx \omega_\lambda$.  The equation for the threshold current
of an $N_r$ times recirculating ERL $I_{\rm th}^{N_r}$ corresponds therefore
to that of the case without recirculation in Eq.~(\ref{eq:simpbbu}),
\begin{equation}
I_{th}^{N_r} = -\frac{2c^2}{e(\frac{R}{Q})_\lambda Q_\lambda
\omega_\lambda}\frac{1}{\sum\!\!\!\!\!\!\!\!\sum\limits_{I>J}
\sin(\omega[t^I-t^J])T^{I\!J}}\ .
\end{equation}

A comparison with Eq.~(\ref{eq:simpbbu}) shows that this current is
smaller than the one-turn ERL by a factor of up to
$\sum\!\!\!\!\!\!\!\!\sum_{I>J} |T^{I\!J}|/|T_{12}|$. This is
in agreement with earlier result presented in
\cite{randbook}. Assuming that all matrix elements are of about equal
magnitude, the threshold current in an $N_r$ times recirculating ERL
is therefore in general smaller by about a factor of
$N_r(2N_r-1)$. This conclusion is consistent with tracking results for
microtrons \cite{microtron} and for two-turn ERL
\cite{twopassmemo}. The scaling in a particular case, however, can be
quite different depending on details of the lattice design,
e.g.~approximate scaling with $N_r$ was reported in \cite{hermin}.

\section{Cavity Misalignments}

In the derivation above, it was assumed that the bunches travel
along the cavities' symmetry axes when the current is below the
threshold for BBU instability. When the cavities are misaligned, the
beam will excite dipole higher order modes even below the threshold and the
trajectory will be disturbed by these modes.

Let us assume that the $i$th cavity is misaligned with respect to the
path adjustment of the $I$th turn by $x_{0i}^I$, leading to the
misalignment vector $\vec x_0$.  The dipole HOMs that are excited by
the beam are now not only due to the beam position fluctuation that is
produced by the HOMs themselves, but additionally due to the cavity
misalignments. The HOM voltages in Eq.~(\ref{eq:homsum}) are therefore
given by
\begin{equation}
V_i(t) = \int_{-\infty}^t
\sum_{I=1}^{N_p}W_i(t-t')I_i^I(t')[x_i^I(t')-x_{0i}^I]dt'\ .
\end{equation}
With the manipulations that led to Eq.~(\ref{eq:hommultsum}) this leads to
\begin{eqnarray}
V_i(t) &=& I_0 t_b \sum_{m=-\infty}^{n(t,t_i^I)}
\sum_{I=1}^{N_p} W_i(t-t_i^I-m t_b)\label{eq:hommultsum2}\\ &\times&
\left[\sum_{J=1}^I\sum_{j=1}^{N_{I\!J}(i-1)} T_{ij}^{I\!J}\frac{e}{c}V_j(m
t_b+t_j^J)-x_{0i}^I\right]\ .\nonumber
\end{eqnarray}
Following the derivation to Eq.~(\ref{eq:matdef}) leads to
\begin{eqnarray}
V_i^L
&=&
I_0t_b\sum_{I=1}^{N_p}w_i(\delta_i(I,L))
e^{i\omega(t_i^I-t_i^L+\delta_i(I,L)t_b)}\nonumber\\
&\times&
\left[\sum_{J=1}^I\sum_{j=1}^{N_{I\!J}(i-1)} T_{ij}^{I\!J}\frac{e}{c}V_j^J
-x_{0i}^I\sum_{n=-\infty}^\infty e^{i\omega n t_b}
\right]\ .
\end{eqnarray}
For all oscillation frequencies that are not multiples of the bunch
repetition frequency, $\omega=2\pi l /t_b$, $l$ is an integer, the term
$x_{0i}^I\sum_{n=-\infty}^\infty e^{i\omega n t_b}$ vanishes so that the
condition for $V_i^L(\omega)$ to be non-zero is the same as for the
BBU instability without misalignments $x_{0i}^I$. Below threshold, the
HOM voltage therefore has the following form:
\begin{equation}
V_i(t)=\sum_{l=-\infty}^\infty V_l e^{i\frac{2\pi}{t_b}l t}\ .
\end{equation}
The voltage seen by bunch $i$ on turn $L$ is therefore given by $\bar
V_i^L=\sum_{l=-\infty}^\infty V_l e^{i\frac{2\pi}{t_b}l t_i^L}$ so
that for the time $t_i^L$, Eq.~(\ref{eq:hommultsum2}) can be written as
\begin{eqnarray}
\bar V_i^L &=& I_0 t_b \sum_{m=-\infty}^{n(t_i^L,t_i^I)}
\sum_{I=1}^{N_p} W_i(t_i^L-t_i^I-m t_b)\\ &\times&
\left[\sum_{J=1}^I\sum_{j=1}^{N_{I\!J}(i-1)} T_{ij}^{I\!J}\frac{e}{c}
\bar V_j^J-x_{0i}^I\right]\ .\nonumber
\end{eqnarray}
After performing the summation over $m$ one obtains
\begin{eqnarray}
\bar V_i^L &=& I_0 t_b
\sum_{I=1}^{N_p} w_i^{\omega=0}(\delta(I,L))\\ &\times&
\left[\sum_{J=1}^I\sum_{j=1}^{N_{I\!J}(i-1)} T_{ij}^{I\!J}\frac{e}{c}
\bar V_j^J-x_{0i}^I\right]\ ,\nonumber
\end{eqnarray}
where the superscript $\omega=0$ means that $w(\delta)$ is computed with
Eq.~(\ref{eq:wsum}) for $\omega=0$.

Comparing with Eqs.~(\ref{eq:matdef}) and (\ref{eq:vmat}) shows that this
can be written as
\begin{equation}
\vec{\bar V}=I_0{\bf W}(0){\bf U}\vec{\bar V}-I_0{\bf W}(0)\vec x_0\ ,
\end{equation}
with ${\bf W}={\bf W}(0)$ and ${\bf U}$ from
Eqs.~(\ref{eq:wdef}) and (\ref{eq:udef}).  The vector of voltages is
therefore given by $\vec{\bar V}=[I_0{\bf WU}-{\bf I}]^{-1}I_0{\bf W}\vec x_0$.
Equation (\ref{eq:xfromv}) can now be used to compute the beams distance from
the cavity center at each turn and each cavity. One obtains
\begin{eqnarray}
x_i^I(t_i^I) - x_{0i}^I &=& \sum_{J=1}^{I} \sum_{j=1}^{N_{I\!J}(i-1)}
T_{ij}^{I\!J} \frac{e}{c} \bar V_j^J- x_{0i}^I\ ,\\ \vec x-\vec x_0
&=& {\bf U} [I_0{\bf WU}-{\bf I}]^{-1}I_0{\bf W}\vec x_0-\vec
x_0\label{eq:xcomp}\\ &=& [I_0{\bf UW}-{\bf I}]^{-1}\vec x_0 \
.\nonumber
\end{eqnarray}

Evaluating this for a single cavity with a single HOM leads to 
\begin{equation}
x-x_0=\frac{x_0}{\frac{e}{c} I_0 t_b T_{12}w^{\omega=0}(\delta)-1}\ .
\end{equation}
For $\delta = \frac{1}{2}$ and with
$\epsilon=\frac{\omega_\lambda}{2Q_\lambda}t_b$ one obtains

\begin{equation}
x-x_0
=
x_0\left[I_0\mathcal{K}T_{12}
\frac{\cosh\frac{\epsilon}{2} \sin\frac{\omega_\lambda t_b}{2}}
{\cosh\epsilon -\cos\omega_\lambda t_b}
-1\right]^{-1}\ .\label{eq:x1comp}
\end{equation}

There is a current $I_0$ at which the denominator becomes $0$ and the
orbit deviation would become very large.  The question arises whether
this current is larger than the BBU threshold $I_{\rm th}$ or smaller, so that
large orbit excursions would present a new kind of instability.

This problem does not only arise for the single HOM case of
Eq.~(\ref{eq:xcomp}) but also for the general case of
Eq.~(\ref{eq:x1comp}). Very large orbit excursions $x$ occur for
currents for which the matrix inverse does not exist.

The inverse matrix to be inverted is $(I_0{\bf WU}-{\bf
1})^{-1}=I_0{\bf A}\,{\rm Diag}[(\lambda_i-1/I_0)^{-1}]\,{\bf A}^{-1}$,
where ${\bf A}$ is the matrix that diagonalizes ${\bf W}(0){\bf U}$. We
therefore see that $1/I_0$ for which the orbit gets very large
is given by the eigenvalues of ${\bf W}(0){\bf U}$. These values are
naturally smaller than $1/I_{\rm th}$, which is the largest
eigenvalue of ${\bf W}(\omega){\bf U}$ that is produced for any frequency
$\omega$.

This proves that the BBU instability always occurs before the orbit
excursion becomes very large.

\section{Tracking Results}
The tracking code \verb+BI+ (stands for beam instability) was
developed to perform studies of beam breakup in recirculating linacs
\cite{bbucode}.  The algorithm models point charge bunch interactions
with HOMs in linacs, taking into account proper time delays between
the cavities, transfer maps, etc., allowing BBU simulations due to
longitudinal, transverse and other higher order modes in a general
linac configuration.

The basic algorithm can be summarized as following. The string of HOMs
that a bunch sees in its lifetime between injection and ejection
points is represented by a list of pointers to the actual
cavities. The proper time delays between cavities is also stored for
each pointer.  E.g.~for $N$ HOMs and $N_p$ passes, the list of
pointers would be $N N_p$ long pointing to $N$ HOMs. This approach
allows one to represent any recirculation configuration without
limitations. As the train of bunches is injected into the structure,
the next instance when any bunch sees any pointer is determined, and
the HOM voltage in the corresponding cavity is updated. Then, this
bunch is pushed to the next pointer where its coordinates are stored,
waiting for its turn in time to be the next bunch going through a
pointer. This way no bunches end up ahead of time precluding a
situation when a bunch sees a cavity with incorrectly updated HOM
fields, i.e. causality is properly realized. Furthermore, the
algorithm is general enough to allow modeling of the longitudinal
instability where timing between different bunches is no longer kept
fixed.  The practical realization of this algorithm is relatively
fast, allowing the tracking of a complete $5\,$GeV ERL with $300$ HOMs
for $0.1\,$ms in less than a minute on an average personal computer.
This duration is sufficient to determine the onset of transverse BBU
instability in most practical cases.

The output of the code contains amplitudes of HOM voltages as a
function of time, which is used to determine the growth rate of the
instability by fitting an exponential. Several successive calls are made
to the tracking unit to determine the threshold.  The length of
the tracking time is estimated from Eq.~(\ref{eq:rate}) based on the
desired accuracy in threshold determination.

\section{Conclusion}
For dipole HOMs we have derived the BBU theory for arbitrary
recirculation times, so that the theory can be applied to ERLs.
The resulting equations have been used to find analytical results for (1)
multiple HOMs in one cavity, (2) two equal HOMs for a one-turn ERL,
and (3) one HOM for a multiple times recirculating ERL.  For (1) the
numerical observation that it often suffices to include only the
strongest of several different HOM was explained and the distance in
frequency was derived for which one can consider two HOMs as different,
for (2) it was shown that two cavities do not cancel each others
instability but that they can be arranged so that they do not add
dangerously, and for (3) was shown that the BBU threshold current for
an $N_r$ times recirculating ERL is roughly up to $N_r(2N_r-1)$ times
smaller than that in a corresponding one-turn ERL.

Furthermore, a simple method to compute the orbit deviations produced
by cavity misalignments has also been introduced. And it is shown that
the BBU instability always occurs before the orbit excursion becomes
very large.

Several comparisons with tracking data verify the applicability of the
theory and of the tracking program.  The conclusions should be useful
in determining and optimizing the maximum current for the many ERLs
that are currently in design and pre-proposal stages worldwide.

\section{Acknowledgments}
We thank Joseph Bisognano and Geoffrey Krafft for careful reading of
the manuscript and for their useful comments.


\begin{thebibliography}{9}

\bibitem{Tigner65_01} M.~Tigner, Nuovo Cimento {\bf 37}, 1228 (1965).

\bibitem{CHESS01_03} S.M.~Gruner, M.~Tigner (eds.), Report No. CHESS 01-003, 2001.

\bibitem{ERL03_12} G.H.~Hoffstaetter {\it et al.}, in
{\it Proceedings of the 2003 Particle Accelerator Conference,
Portland, OR} (IEEE, Piscataway, NJ, 2003), pp.~192-194.

\bibitem{Benzvi01_02} I.~Ben-Zvi {\it et al.}, in 
{\it Proceedings of the 2001 Particle Accelerator Conference,
Chicago, IL} (IEEE, Piscataway, NJ, 2001), pp.~350-352.

\bibitem{Pool03_01} M.W.~Poole {\it et al.}, in 
{\it Proceedings of the 2003 Particle Accelerator Conference,
Portland, OR} (IEEE, Piscataway, NJ, 2003), pp.~189-191.

\bibitem{Benson01_01} S.V.~Benson {\it et al.}, in 
{\it Proceedings of the 2001 Particle Accelerator Conference,
Chicago, IL} (IEEE, Piscataway, NJ, 2001), pp.~249-252.

\bibitem{Sawamura03_02} M.~Sawamura {\it et al.}, in 
{\it Proceedings of the 2003 Particle Accelerator Conference,
Portland, OR} (IEEE, Piscataway, NJ, 2003), pp.~3446-3448.

\bibitem{Berkaev02_01} D.E.~Berkaev {\it et al.}, in
{\it Proceedings of the 2002 European Particle Accelerator Conference,
Paris, France} (CERN, Geneva, 2002), pp.~724-726.

\bibitem{Kulipanov98_01} G.N.~Kulipanov, A.N.~Skrinsky,
N.A.~Vinokurov, J. Synchrotron Rad. {\bf 5}, 176 (1998).

\bibitem{Suwada02_01} T.~Suwada {\it et al.}, in 
{\it Proceedings of the 2002 ICFA Beam Dynamics Workshop
on Future Light Sources, Japan}.

\bibitem{Merminga02_01} L.~Merminga {\it et al.}, in 
{\it Proceedings of the 2002 European Particle Accelerator Conference,
Paris, France}, (CERN, Geneva, 2002), pp.~203-205.

\bibitem{Benzvi03_01} I.~Ben-Zvi {\it et al.}, in
{\it Proceedings of the 2003 Particle Accelerator Conference,
Portland, OR} (IEEE, Piscataway, NJ, 2003), pp.~39-41.

\bibitem{Bisognano87_01} J.J.~Bisognano, R.L.~Gluckstern, 
in {\it Proceedings of the 1987 Particle Accelerator Conference,
Washington, DC} (IEEE Catalog No.~87CH2387-9), pp.~1078-1080.

\bibitem{Krafft87_01} G.A.~Krafft, J.J.~Bisognano, in {\it Proceedings of the
1987 Particle Accelerator Conference, Washington, DC}
(IEEE Catalog No.~87CH2387-9), pp.~1356-1358. 

\bibitem{Beard03_01} K.~Beard, L.~Merminga, B.C.~Yunn, in 
{\it Proceedings of the 2003 Particle Accelerator Conference,
Portland, OR} (IEEE, Piscataway, NJ, 2003), pp.~332-334.

\bibitem{Merminga01_01} L.~Merminga, I.E.~Campisi, D.R.~Douglas,
G.A.~Krafft, J.~Preble, B.C.~Yunn, in 
{\it Proceedings of the 2001 Particle Accelerator Conference,
Chicago, IL} (IEEE, Piscataway, NJ, 2001), pp.~173-175.

\bibitem{Bisognano86_01} J.J.~Bisognano, G.A.~Krafft, in {\it Proceedings
of the 1986 Linear Accelerator Conference, Stanford, CA} (SLAC-303), 
pp.~452-454.

\bibitem{krafft89} G.A.~Krafft, J.J.~Bisognano, S.~Laubach,
unpublished, 1987.

\bibitem{Sereno94} N.S.R.~Sereno, Ph.D.~Dissertation,
University of Illinois, 1994

\bibitem{Yunn91_01} B.C.~Yunn, in {\it Proceedings of the 1991
Particle Accelerator Conference, San Francisco, CA}
(IEEE Catalog No.~91CH3038-7), pp.~1785-1787.

\bibitem{randbook} R.E.~Rand, {\it Recirculating electron accelerators}
(Harwood Academic Publishers, New York, 1984), Section 9.5.

\bibitem{microtron} L.~Merminga, B.C.~Yunn, JLAB Technical Report No.
TN-97-032, 1997

\bibitem{twopassmemo} I.V.~Bazarov, LEPP Technical Report No. ERL 02-4, 2002,
\verb+http://lepp.cornell.edu/public/ERL/+

\bibitem{hermin} H.~Herminghaus, H.~Euteneuer,
Nuclear Instruments and Methods {\bf 163}, 299 (1979).

\bibitem{bbucode} Available at
\verb+http://lepp.cornell.edu/~ib38/+

\end{thebibliography}
\end{document}